%% file: bd-HJHJ-twocol.tex
\documentclass[twocolumn,aps,superscriptaddress]{revtex4}

\makeatletter
\def\@dotsep{4.5}
\makeatother

\usepackage{natbib}
\usepackage{definitions}
\usepackage{localdefs}
\usepackage{amsmath}
\usepackage{amssymb}
\usepackage{graphics}
\usepackage{amsbsy}
\usepackage{bm}

\begin{document}
\date{\today}

\title{Effect of small particles on the near-wall dynamics of a large
particle in a highly bidisperse colloidal solution}

\author{S. Bhattacharya}
\affiliation{Department of Mechanical Engineering, Texas Tech University,
Lubbock, TX-79409}

\author{J. Blawzdziewicz}
\affiliation{Department of Mechanical Engineering, Yale University, New Haven,
CT-06520}

\begin{abstract}

We consider the hydrodynamic effect of small particles on the dynamics of a much larger particle moving normal to a planar wall in a highly bidisperse dilute colloidal suspension of spheres.  The gap $h_0$ between the large particle and the wall is assumed to be comparable to the diameter $2a$ of the smaller particles so there is a length-scale separation between the gap width $h_0$ and the radius of the large particle $b\gg h_0$. We use this length-scale separation to develop a new lubrication theory which takes into account the presence of the smaller particles in the space between the larger particle and the wall.  The hydrodynamic effect of the small particles on the motion of the large particle is characterized by the short time (or high frequency) resistance coefficient.  We find that for small particle--wall separations $h_0$, the resistance coefficient tends to the asymptotic value corresponding to the large particle moving in a clear suspending fluid.  For $h_0\gg a$, the resistance coefficient approaches the lubrication value corresponding to a particle moving in a fluid with the effective viscosity given by the Einstein formula.

\end{abstract}

\maketitle

\section{Introduction}
\label{Introduction}

Polydisperse suspensions have a significant presence in many important
industrial and natural systems. Examples include dairy products, polymeric
gels and nearly all complex fluids in biological organisms.  Since the
dynamics of colloidal systems is often influenced by their confining
boundaries \cite[][]{Oetama-Walz:2006, Oetama-Walz:2006a}, the confining effects 
are important in many emerging applications, e.g., in self-assembly technology 
involving colloidal crystals \cite[][]{Subramanian-Manoharan-Thorne-Pine:1999,
Seelig-Tang-Yamilov-Cao-Chang:2002} and in chemical deposition processes on patterned wall
\cite[][]{Lin-Crocker-Prasad-Schofield-Weitz-Lubensky-Yodh:2000}. Hence, it is
necessary to understand how polydisperse suspensions behave in the proximity
of a bounding surface.

The equilibrium state of confined systems can be fully determined by the
particle--wall and interparticle interaction potentials, whereas a description
of non-equilibrium phenomena is much more complex.  In this paper we study the
non-equilibrium dynamics of a wall-bounded bidisperse solution where a large
colloidal particle moves in the direction normal to the wall in the presence
of smaller particles. Our objective is to find the correction to the
near-contact hydrodynamic force on the large particle due to the effect of the
smaller particles in the low-concentration regime.

One way to tackle this problem is to treat the suspension in the gap between
the wall and the large particle as a fluid with the effective viscosity given
by the Einstein's formula \cite[][]{Einstein:1906}.  This approach, however,
is valid only if the dimension of the small particles is much smaller than the
characteristic flow length scale which is the minimum distance of separation
between the surfaces of the wall and the large particle.  When the separating
distance is comparable to the dimension of the smaller species, a more
detailed analysis is necessary to obtain the proper correction to the
hydrodynamic friction.  Our main purpose is to evaluate this correction in the
short time (or high frequency) regime, where the influence of the small
particles on the large particle motion results entirely from hydrodynamic
effects.

This article is organized in the following way. In section \ref{Highly
asymmetric bidisperse colloidal suspension}, we define our system and discuss
assumptions involved in our calculations. In section \ref{Lubrication solution
in near-contact region}, we develop a lubrication theory for near-contact
approach of the large particle towards the wall and use this theory to derive
expressions for modified lubrication pressure and hydrodynamic force on the
large particle in the presence of small particles.  In section
\ref{Ensemble-averaged force density}, we express the dipolar source density
in the lubrication equations in terms of the hydrodynamic forces induced on
the surfaces of small particles.  In section \ref{Effective resistance force}
the results of our theory are used to determine the resistance force acting on
the large particle.  Our conclusions are drawn in section \ref{Conclusion}.

\input{geo}
\section{Highly asymmetric bidisperse colloidal suspension}
\label{Highly asymmetric bidisperse colloidal suspension}

We consider a bidisperse colloidal suspension near a solid planar surface.
Particles of both species are assumed to be spherical, with high asymmetry in
size. We describe the influence of the smaller particles on the motion of a
single larger particle in the direction normal to the wall. Our analysis is
focused on the effect of the smaller species when the separation between the
larger particle and the wall is comparable to the diameter of the smaller
particles.

We concentrate on the short-time (or high-frequency) regime; the
effect of Brownian motion on the system dynamics can therefore be
neglected on the assumption that the deviation of the particle
distribution from equilibrium is small.  Under these conditions, the
smaller particles affect the motion of a large particle only by
influencing the velocity and the pressure fields in the suspending
fluid. In order to determine these hydrodynamic fields, we first
analyze the flow in the presence of a single small particle in the gap
between the wall and the large particle.  The cumulative effect of
small particles in a dilute suspension is obtained by simple
statistical averaging over the position of the smaller particles. This
approach is valid in the concentration range where the suspension is
dilute enough to neglect mutual interactions between smaller particles
and dense enough for meaningful statistical averaging.

\subsection{System definition}
\label{Geometry}

The geometry of our system is represented in Fig.\ \ref{geo-fig}. The
small and the large particles have spherical hydrodynamic cores of
radii $a$ and $b$, where $a\ll b$.  In typical highly asymmetric
bidisperse colloidal mixtures $b/a$ ranges from $10^2$ to $10^3$.

A rigid wall is located at $z=0$, and the center of the large sphere
is on the $z$-axis. Therefore, the surface of this particle can be
described by $z=h(\rho)$ where $\rho$ is the distance from the axis of
symmetry ($z$-axis). The minimum distance of separation between the
particle surface and the wall is $h(0)=h_{0}$. We consider $h_{0}\sim
2a$.

A repulsive screened electrostatic potential between the particles is
modeled by additive excluded shells surrounding hydrodynamic core of
each particle.  The thicknesses of the excluded shells for the small
and the large spheres are $\delta_{a}$ and $\delta_{b}$,
respectively. The thickness of the excluded layer associated with the
wall is denoted by $\delta_{w}$. It is assumed that
$\delta_{a},\delta_{b},\delta_{w}\ll b$.  

The interaction potential affects the short-time dynamics of the system only
through its influence on the equilibrium particle distribution---the effective
resistance coefficient is obtained as the equilibrium average of the
microscopic stresses produced by the particle motion.  The results for a
general interaction potential can thus be obtained by evaluating an
appropriate equilibrium average of the hydrodynamic functions derived in the
following sections.  Our explicit numerical results for the excluded-shell
potential illustrate the essential features of the near-wall dynamics of large
particles in a highly bidisperse colloidal suspension.

\subsection{Equation for the flow field}

In our calculations, velocity of the large colloidal particle is denoted as
$V^{\rm P}$ which is in the direction perpendicular to the wall. In this paper, 
all of our results correspond to the case where the large particle approaches 
to the wall---the reversal in this motion can be accounted for by simple reversal 
of computed velocity fields and pressure gradients. The velocity $V^{\rm P}$ 
is small so that the creeping flow approximation is valid. Hence, the hydrodynamic 
fields are governed by the Stokes equations,
\begin{equation}
\label{Stokes eq. wo source}
-\bnabla P+\eta\nabla^{2}{\bf u}=0, \qquad\qquad \bnabla\cdot{\bf u}=0,
\end{equation}
where ${\bf u}$ is the velocity field, $P$ is the pressure field, and
$\eta$ is the viscosity. No-slip boundary condition is assumed at the
wall and at the surfaces of all particles.

The hydrodynamic effect of the small particles on the surrounding fluid can be
described in terms of the induced force density ${\bf F}$ on surface of the
small spheres. Such force density imposes a discontinuity in the gradient of
the velocity field at the particle--fluid interface. This discontinuity is
such that we can consider the volume inside the particle as an extension of
the fluid domain---the fluid in this region undergoes rigid-body motion
corresponding to that of the rigid particle. Therefore, the entire domain
(both inside and outside the particle) can be treated as a continuous fluid
medium. As a result, the flow field described by Eq.\ \eqref{Stokes eq. wo
source} in the presence of the smaller particles is equivalent to the flow
field which obeys the inhomogeneous Stokes equations
\begin{equation}
\label{Stokes eq.}
-\bnabla P+\eta\nabla^{2}{\bf u}={\bf F}, \qquad\qquad \bnabla\cdot{\bf u}=0.
\end{equation}
Hence, if we can determine ${\bf F}$ and solve Eq.\ \eqref{Stokes
eq.}, we can find the effect of the smaller species on the velocity
and the pressure fields. Accordingly, in the first part of our
analysis, we solve Eq.\ \eqref{Stokes eq.} to find $P$ and ${\bf u}$
in the gap between the large particle and the wall for an arbitrary
induced-force distribution ${\bf F}$.  This analysis is presented in
section \ref{Lubrication solution in near-contact region}.

\subsection{Statistically averaged force density}

In order to quantify the cumulative influence of smaller species on
the hydrodynamic resistance of the large particle, we solve an
ensemble-averaged flow equation (where the average is taken over all
possible positions of the small particles). This reduced description
requires the averaging of the force density ${\bf F}$.  Such averaging
is valid if two conditions are satisfied. Firstly, there should be a
sufficient number of small particles in the gap between the large
particle and the wall so that the statistical description is
physically meaningful. Secondly, the small particles should be well
separated and mutually non-interacting.  These two conditions are
satisfied for
\begin{equation}
\label{semidilute}
h_{0}\ll C^{-\frac{1}{3}}\ll\sqrt{bh_{0}},
\end{equation}
with $C$ being the number density of the small particles. 
Under these conditions, the statistically averaged force distribution
can be evaluated from the cumulative effect of mutually noninteracting
individual small particles. 

\section{Lubrication solution in the near-contact region}
\label{Lubrication solution in near-contact region}

In our hydrodynamic calculations, we exploit the fact that for
near-contact motion of the large particle the characteristic length
scale along the wall is much greater than the characteristic length
scale along the perpendicular direction $z$. As a result of this
separation of length scales, we can apply lubrication theory to
determine the flow and the pressure fields.

\subsection{Lubrication expansion}

The predominant hydrodynamic effect on the large particle near a wall
comes from the flow field in the near-contact lubrication region. In
this lubrication zone, the flow can be locally approximated as a flow
between two rigid parallel planes. The extent of the lubrication
region is defined by the lubrication length $l$, which is the
characteristic length in the direction along the wall:
\begin{equation}
\label{sep of length scale}
l=\sqrt{2ab}\sim\sqrt{bh_{0}}\gg h_{0}\sim 2a.
\end{equation} 
The corresponding characteristic length for the direction
perpendicular to the wall is the slowly varying gap width $h\sim 2a\ll
l$.   The small parameter that characterizes length-scale separation
in the two directions is
\begin{equation}
\label{size ratio sep of length scale}
\lambda=2a/l=\sqrt{2a/b}.
\end{equation} 

In our earlier paper \cite[][]{Bhattacharya-Blawzdziewicz-Wajnryb:2006}, we
have shown that the scattered flow from a sphere in a slit pore saturates
exponentially to an asymptotic Hele--Shaw velocity field at large distances
from the sphere.  In the asymptotic regime, the scattered flow thus assumes
the form of a two-dimensional pressure-driven flow, with a harmonic pressure
distribution which is independent of the transverse coordinate $z$.

There are two ways in which the flow scattered from the small
particles contributes to the hydrodynamic normal force acting on the
large particle.  First, there is the near-field behavior of the
scattered field, which produces stresses confined to a localized
region of size defined by the length $h\sim a$.  Second, there is also the
far-field lubrication pressure.

To estimate the hydrodynamic force resulting from the near-field interactions,
we note that there are $N\sim C l^2a$ small particles in the lubrication
region, and each particle yields an $O(\eta a V^{\textrm{P}})$ force
contribution.  We thus find that the cumulative effect of the near-field
stresses on the total resistance force acting on the large particle is of the
order $O(\phi\eta b V^{\textrm{P}})$, where $\phi$ is the volume fraction of
the small particles.  This force is thus smaller by the order $O(a/b)$ than
the lubrication resistance force
\begin{equation}
\label{brenner}
F_{0}=6\pi\eta V^{\rm P}b^{2}/h_{0}
\end{equation}
resulting from the presence of the suspending fluid
\cite[][]{Cox-Brenner:1967}.

The second small-particle contribution to the resistance force originates from
the cumulative effect of the far-field pressure associated with the flow
scattered from the particles.  The far-field pressure decays slowly at large
distances, and thus it acts over the whole lubrication area.  As shown below,
the far-field pressure produces an $O(\phi F_{0})$ contribution to the
resistance force.  This contribution is thus the dominant correction to the
hydrodynamic resistance.

To find the far-field contribution of the scattered flow, we rescale the
lateral coordinates ${\bf s}$ (parallel to the wall) and transverse coordinate
$z$ (across the gap) by the corresponding lengthscales $l$ and $2a$,
\begin{equation}
\label{rescaling of lateral and transvere dimensions}
{\bf s}\rightarrow {\bf s}/l,\qquad z\rightarrow z/2a.
\end{equation}
Accordingly, the Stokes equations \eqref{Stokes eq.} are
transformed into the rescaled form,
\begin{subequations}
\label{mod. Stokes eq.}
\begin{equation}
\label{mod. Stokes eq. z}
-\frac{1}{2a}\frac{\partial P}{\partial z}
   +\frac{\eta}{4a^{2}}\left(
          \frac{\partial^{2}u_{z}}{\partial z^{2}}
          +\lambda^{2}\nabla_{s}^{2}u_{z}
   \right)=F_{z},
\end{equation}
\begin{equation}
\label{mod. Stokes eq. horz.}
-\frac{\lambda}{2a}\bnabla_{s} P
   +\frac{\eta}{4a^{2}}\left(
          \frac{\partial^{2} \bu_{s}}{\partial z^{2}}
          +\lambda^2\nabla_{s}^{2}{\bf u}_{s}
   \right)={\bf F}_{s}, 
\end{equation}
\begin{equation}
\label{mod. Stokes eq. div}
\lambda\bnabla_{s}\cdot{\bf u}_{s}+\frac{\partial u_{z}}{\partial z}=0,
\end{equation}
\end{subequations}
where subscripts $s$ and $z$ denote the directions parallel and normal
to the wall, respectively.  

From now on, all coordinates and position vectors will be normalized
by the corresponding characteristic lengths. To be consistent with
this non-dimensional formulation, we introduce the following
dimensionless parameters:
\begin{subequations}
\label{ratios}
\begin{equation}
\label{three ratios}
\bar{b}=b/2a=1/\lambda^{2},\quad{\bar h}_{0}=h_{0}/2a,
\quad\bar{h}=h/2a,
\end{equation}
and
\begin{equation}
\label{dimensionless delta}
\quad{\bar\delta}_{i}=\delta_{i}/2a.
\end{equation} 
\end{subequations}
The subscript $i$ in Eq.\ \eqref{dimensionless delta} can be either
$a$ (the small spheres) or $b$ (the large sphere) or $w$ (the wall) to
indicate the excluded-volume range due to corresponding repulsive
potentials. For simplicity, in our calculations, we take
\begin{equation}
\label{excld. vols}
{\bar\delta}_{a}={\bar\delta}_{w}={\bar\delta}_{b}=\bar{\delta}/2
\end{equation}
and use $\bar{\delta}$ as the only parameter which accounts for the
excluded volume. The behavior of more general systems is similar to
the one where Eq.\ \eqref{excld. vols} is assumed.

The rescaled Stokes equations \eqref{mod. Stokes eq.} can be solved
asymptotically by scaling the velocity and the pressure fields with
proper dimensional quantities and expanding these fields in the small
parameter $\lambda$,
\begin{subequations}
\label{expansion of velocity and prs.}
\begin{equation}
\label{expansion of prs.}
P=\frac{V^{\rm P}\eta}{2a}\,\frac{1}{\lambda^{2}}\sum_{i=0} \lambda^{2i}P_{i},
\end{equation}
\begin{equation}
\label{expansion of velocity s}
{\bf u}_{s}=V^{\rm P}\frac{1}{\lambda}\sum_{i=0} \lambda^{2i}{\bf u}_{si}, 
\end{equation}
\begin{equation}
\label{expansion of velocity}
u_{z}=V^{\rm P}\sum_{i=0} \lambda^{2i}u_{zi}.
\end{equation}
\end{subequations}
The different leading-order powers of $\lambda$ in the expansion of $P$, ${\bf
u}_{s}$ and $u_{z}$ stem from the different orders of various terms in Eqs.\
\eqref{mod. Stokes eq.}.  By inserting expansion \eqref{expansion of velocity
and prs.} into \eqref{mod. Stokes eq.}  and collecting terms with the same
power of $\lambda$, a hierarchy of equations corresponding to each order of
$\lambda$ can be obtained. We are particularly interested in the zeroth order
terms which satisfy the following leading-order lubrication equations:
\begin{subequations}
\label{lubrication eq.}
\begin{equation}
\label{lubrication eq. z-momentum}
\frac{\partial P_{0}}{\partial z}=0,
\end{equation}
\begin{equation}
\label{lubrication eq. s-momentum}
-\bnabla_{s}P_{0}+\frac{\partial^{2}{\bf u}_{s0}}{\partial z^{2}}={\bf f}_{s},
\end{equation}
\begin{equation}
\label{lubrication eq. cont.}
\frac{\partial u_{z0}}{\partial z}+\bnabla_{s}\cdot{\bf u}_{s0}=0,
\end{equation}
\end{subequations}
where
\begin{equation}
\label{non-dim F}
{\bf f}_{s}=\frac{4a^{2}\lambda}{\eta V^{\rm P}}\,{\bf F}_{s}.
\end{equation}
The leading-order hydrodynamic effect is obtained by solving the
lubrication equations \eqref{lubrication eq.}.

\subsection{Leading order pressure and friction}

The momentum equation in the $z$-direction \eqref{lubrication
eq. z-momentum} implies that $P_{0}$ is independent of $z$. Using this
fact along with the momentum equation in the lateral direction
\eqref{lubrication eq. s-momentum} and the no-slip boundary conditions
at the solid surfaces, ${\bf u}_{s0}$ can be determined in terms of
the pressure field $P_0$,
\begin{widetext}
\begin{equation}
\label{lubrication sol. s-momentum}
{\bf u}_{s0}=\frac{1}{2}z(z-\bar{h})\bnabla_{s}P_{0}
  +\int_{0}^{z}\int_{0}^{z'}{\bf f}_{s}dz''dz'
  -\frac{z}{\bar{h}}\int_{0}^{\bar{h}}\int_{0}^{z'}{\bf f}_{s}dz''dz'.
\end{equation}
\end{widetext}
Then, the velocity perpendicular to the wall can be evaluated by
integrating the continuity equation \eqref{lubrication eq. cont.}
\begin{equation}
\label{lubrication sol. cont.}
u_{z0}=-\int_{0}^{z}\bnabla_{s}\cdot{\bf u}_{s0}\,dz',
\end{equation}
and using the boundary condition at the surface of the large particle
\begin{equation}
\label{uz0 at surface of the large particle}
u_{z0}|_{z=h}=-1.
\end{equation}
We combine Eqs.\ \eqref{lubrication sol. s-momentum}--\eqref{uz0 at
surface of the large particle} to obtain the lubrication equation for
the zeroth order pressure field,
\begin{equation}
\label{tot lubrication prs. eq.}
\bnabla_{s}\cdot(\bar{h}^{3}\bnabla_{s}P_{0}+{\bf b}_{s})=-12.
\end{equation}
The source term ${\bf b}_{s}$ in Eq.\ \eqref{tot lubrication prs. eq.}
is given by
\begin{equation}
\label{lubrication force distribution}
{\bf b}_{s}(\mbox{\boldmath{$\rho$}})=\hat{\sf B}\,\,{\bf f}_s({\bf r}),
\end{equation}
where
\begin{widetext}
\begin{equation}
\label{B operator}
\hat{\sf B}\,{\bf f}_{s}
    =-12\int_{0}^{\bar{h}}
\left(
    \int_{0}^{z}\int_{0}^{z'}{\bf f}_{s}dz''dz'
    -\frac{z}{\bar{h}}\int_{0}^{\bar{h}}\int_{0}^{z'}{\bf f}_{s}dz''dz'
\right)\,dz.
\end{equation}
\end{widetext}

In the absence of smaller species (i.e., when ${\bf f}_{s}=0$), well
known lubrication equation for the pressure field in a slowly varying
gap can be recovered from Eq.\ \eqref{tot lubrication prs. eq.},
\begin{equation}
\label{real lubrication prs. eq.}
\bnabla_{s}\cdot(\bar{h}^{3}\bnabla_{s}P_{L})=-12.
\end{equation}
The presence of the smaller spheres produces a pressure correction  
\begin{equation}
\label{perturbation corrective pressure field}
p_{0}=P_{0}-P_{L}
\end{equation}
to the non-dimensional lubrication pressure $P_{L}$. The perturbation
pressure field $p_{0}$ satisfies the lubrication equation
\begin{equation}
\label{lubrication prs. eq.}
\bnabla_{s}\cdot(\bar{h}^{3}\bnabla_{s}p_{0}+{\bf b}_{s})=0,
\end{equation}
which is obtained by combining Eqs.\eqref{tot lubrication prs. eq.},
\eqref{real lubrication prs. eq.} and \eqref{perturbation corrective
pressure field}. 

 The function ${\bf b}_{s}$ in the above equation plays a key role in
 our theory.  It can be interpreted as the dipolar source density for
 the far-field lubrication pressure produced by the induced-force
 distribution $\bF_s$.

The correction in friction of the large particle due to the presence
of the small spheres has contributions from both perturbation pressure
$p_{0}$ and viscous stresses corresponding to the leading order
velocity field ${\bf u}_{0}$. However, comparing the orders of
$\lambda$ in Eq.\ \eqref{expansion of velocity and prs.} we conclude
that the second contribution is negligible compared to the first
one. Thus, the leading order correction in $\lambda$ to the normal
hydrodynamic force acting on the large particle in a solution of small
particles stems from the leading-order pressure correction $p_{0}$. As
a result, the correction in hydrodynamic force due to the presence of
the smaller species can be expressed as
\begin{equation}
\label{leading order force and pressure}
F^{\rm c}_{0}=\frac{\eta b^{2}}{2a}\,\int
p_{0}\,d^{2}\mbox{\boldmath{$\rho$}},
\end{equation}
where $p_{0}$ is assumed to vanish far away from the contact point.

\section{Dipolar source  density $\bb_s$}
\label{Ensemble-averaged force density}

To evaluate the resistance force $F^{\rm c}_{0}$ from Eq.\
\eqref{leading order force and pressure}, we first need to calculate
the lubrication pressure field $p_{0}$.  The pressure can be
determined from the lubrication equation \eqref{lubrication prs. eq.},
provided that the dipolar source term \eqref{lubrication force
distribution} is known. In this section, we focus on the key step
where we compute the source term $\bb_s$ averaged over the positions
of small particles.

\subsection{Force singularities}
\label{Force singularities}

The induced force density at the surface of the $i$-th sphere is equivalent to
a point-force singularity distribution ${\bf F}_{i}$ at the sphere
center. This description implies that the induced force density is considered
at the surface of an infinitely small hypothetical sphere situated at the
center of the $i$-th particle.  However, the strength of the force density is
modified in such a way that only the flow field inside the particle is
altered, and the outside flow remains unchanged.  Accordingly, the force
density for an arbitrary particle $i$ is
\begin{equation}
\label{single sphere force distribution}
{\bf F}_{i}=\frac{\eta V^{\rm P}}{4a^{2}\lambda}
   \sum_{lm\sigma} 
   f_{lm\sigma}\mbox{\boldmath{$\delta$}}_{lm\sigma}(\br-\br_i),
\end{equation}
where $\br_i$ denotes the position of the center of the particle.  The
coefficients $f_{lm\sigma}$ represent the non-dimensional strength of
the force singularity and are referred to as the multipolar
moments. The force singularities
$\mbox{\boldmath{$\delta$}}_{lm\sigma}$ can be expressed in terms of
singular spherical basis solutions of the Stokes equations,
\begin{equation}
\label{Stokes eq. with singular source}
\nabla^2 {\bf v}^{-}_{lm\sigma}-\bnabla p^{-}_{lm\sigma}
   =\mbox{\boldmath{$\delta$}}_{lm\sigma}.
\end{equation}
The singular spherical basis solutions ${\bf v}^{-}_{lm\sigma}$ and
pressure solutions $p^{-}_{lm\sigma}$ are defined in
\cite[][]{Bhattacharya-Blawzdziewicz-Wajnryb:2005,
Bhattacharya-Blawzdziewicz-Wajnryb:2005a}. These fields satisfy the
homogeneous Stokes equation at every point in the flow domain except
at the particle center where they are singular and correspond to the
point-force singularity $\mbox{\boldmath{$\delta$}}_{lm\sigma}$.

By considering the cumulative effect from all the spheres,
we can compute the statistically averaged normalized force
distribution:
\begin{equation}
\label{stat. av. force distribution nd}
{\bf F}({\bf r})
   =\frac{\eta V^{\rm P}}{4a^{2}\lambda}\,\frac{3b}{\pi a}\,
   \int\phi({\bf r}_{0})\sum_{lm\sigma}f_{lm\sigma}({\bf r}_{0})
   \boldsymbol{\delta}_{lm\sigma}({\bf r}-{\bf r}_{0})d^{3}{\bf r}_{0},
\end{equation}
where the volume fraction of small particles $\phi$ is given by
\begin{equation}
\label{vol. frac.}
\phi=\frac{4\pi}{3}a^{3}C.
\end{equation}
The force distribution \eqref{stat. av. force distribution nd}
corresponds to the average induced-force density in the
ensemble-averaged version of the Stokes equation \eqref{Stokes eq.}.
Equation \eqref{stat. av. force distribution nd} is valid for an
arbitrary distribution of small particles $\phi$.  In our present
application the volume fraction is uniform in the available space
outside the excluded-volume shells.

\subsection{Local parallel-wall geometry}
\label{Local parallel-wall geometry}

In order to evaluate the source term ${\bf F}({\bf r})$, the
multipolar moments $f_{lm\sigma}$ need to be computed.  We obtain
these multipolar force distributions by assuming that at the length
scale $a\ll b$, the region confined by the large particle and the wall
can be locally approximated by the geometry of a channel bounded by
two infinite parallel planar walls.  

For such a geometry, the multipolar moments $f_{lm\sigma}$ can be
evaluated with high accuracy either using a multiple-scattering
technique \cite{Bhattacharya-Blawzdziewicz:2002} or a
Cartesian-representation algorithm developed in our recent papers
\cite[][]{Bhattacharya-Blawzdziewicz-Wajnryb:2005,
Bhattacharya-Blawzdziewicz-Wajnryb:2005a,
Bhattacharya-Blawzdziewicz-Wajnryb:2006a}.   

The force multipoles $f_{lm\sigma}({\bf r}_{0})$ are induced as a result of
the interaction between the isolated freely moving small sphere situated at
${\bf r}_{0}$ and the incident horizontal lubrication flow field ${\bf
u}_{Ls}$ created by motion of the large particle,
\begin{equation}
\label{ul ge}
\frac{\partial^{2}{\bf u}_{Ls}}{\partial z^{2}}=\bnabla_{s}P_{L}.
\end{equation}
The lubrication pressure $P_{L}$ satisfies Eq.\ \eqref{real
lubrication prs. eq.}, so that
\begin{equation}
\label{lub prs sol. fnl}
\bnabla_{s}P_{L}=-\frac{6\rho}{\bar{h}^{3}}\hat{\bf e}_{\rho}.
\end{equation}
From Eqs.\ \eqref{ul ge} and \eqref{lub prs sol. fnl}, we find ${\bf
u}_{Ls}$
\begin{equation}
\label{ul}
{\bf u}_{Ls}
  =4\,A_{L}\,\frac{z}{\bar{h}}
  \left(
     1-\frac{z}{\bar{h}}
  \right)
  \hat{\bf e}_{\rho},
\end{equation}
where
\begin{equation}
\label{al}
A_{L}(\rho)=\frac{3\rho}{4\bar{h}}.
\end{equation} 
The amplitude $A_{L}$ depends on the radial position because both the
volume flux and height of the gap are functions of $\rho$.

At the length scale of the small particle diameter, the parabolic
lubrication flow \eqref{ul} can be treated as an external flow coming
from infinity in a parallel-wall channel.  It is convenient to choose
a local coordinate system ($x', y', z'$) with the origin at the center
of the small sphere and the $x'$-axis coinciding with the radial
direction $\hat{\bf e}_{\rho}$.  Hence, the direction of the incident
parabolic flow is along the local $x'$ coordinate.

There are three immediate simplifications in the description of $f_{lm\sigma}$
for a freely suspended sphere in a parabolic flow. Firstly, the multipolar
moments $f_{1\pm 10}$ and $f_{1\pm 11}$ correspond to a Stokeslet and a
rotlet, respectively, and therefore they vanish for a force-free and
torque-free particle. Secondly, for a parabolic Poiseuille flow (like ${\bf
u}_{Ls}$), the multipolar moments are non-zero only for $m=\pm 1$. Thirdly,
our choice of the local coordinates implies that multipolar strength for $m$
is real and is the same as the corresponding value for $-m$.  In the following
section we evaluate the relevant non-zero coefficients $f_{lm\sigma}$ as a
function of the gap width and position of the small particle, and we compute
the pressure source term.

\subsection{The pressure source term}
\label{The pressure source term}

In order to compute $F^{\rm c}_{0}$ from the integral \eqref{leading
order force and pressure}, we need to determine $p_{0}$ in Eq.\
\eqref{lubrication prs. eq.}, which involves the dipolar pressure
source term ${\bf b}_{s}$. We find ${\bf b}_{s}$ from ${\bf f}_s$ by
using Eq.\ \eqref{lubrication force distribution}. Combining Eqs.\
\eqref{non-dim F} and \eqref{stat. av. force distribution nd} along
with the proper non-dimensional scaling, we express ${\bf f}_s$ in the
following form:
\begin{widetext}
\begin{equation}
\label{stat. av. force distribution}
{\bf f}_s({\bf r})=\frac{3b}{\pi a}\int\phi({\bf r}_{0})\sum_{lm\sigma}f_{lm\sigma}({\bf r}_{0})({\bf I}-\hat{\bf e}_z\hat{\bf e}_z)\cdot\mbox{\boldmath{$\delta$}}_{lm\sigma}({\bf r}-{\bf r}_{0})d^{3}{\bf r}_{0},
\end{equation}
\end{widetext}
where ${\bf I}-\hat{\bf e}_z\hat{\bf e}_z$ is the projection tensor on
$x$--$y$ plane. Changing the order of the integrals in Eqs.\
\eqref{lubrication force distribution} and \eqref{stat. av. force
distribution} we obtain ${\bf b}_{s}$:
\begin{equation}
\label{stat. av. bs}
{\bf b}_{s}(\mbox{\boldmath{$\rho$}})=\frac{3 b}{\pi a}\int\phi({\bf r}_{0})\sum_{lm\sigma}f_{lm\sigma}({\bf r}_{0}){\bf b}_{lm\sigma}(\mbox{\boldmath{$\rho$}}-\mbox{\boldmath{$\rho$}}_{0};z_{0})d^{3}{\bf r}_{0},
\end{equation}
where
\begin{equation}
\label{b l m sigma}
{\bf b}_{lm\sigma}(\mbox{\boldmath{$\rho$}}-\mbox{\boldmath{$\rho$}}_{0};z_{0})=({\bf I}-\hat{\bf e}_z\hat{\bf e}_z)\cdot\hat{\sf B}\,\,\mbox{\boldmath{$\delta$}}_{lm\sigma}({\bf r}-{\bf r}_{0}).
\end{equation}
It can be shown that ${\bf b}_{lm\sigma}$ is non-zero only when the
following condition is satisfied
\cite[][]{Bhattacharya-Blawzdziewicz-Wajnryb:2006}
\begin{equation}
\label{parameters of elements contributing to leading order}
l+\sigma-|m|\le 2.
\end{equation}
Due to the symmetry of the problem (a parabolic flow in a slit pore), only the
terms with $m=\pm 1$ are relevant for our present analysis.

In Appendix \ref{derv. liron mochon} we list the expressions for
$\mbox{\boldmath{$\delta$}}_{lm\sigma}$ with $m=\pm 1$ and derive the
corresponding formulas for the source terms ${\bf b}_{lm\sigma}$, which are
obtained by applying the operator $\hat{\sf B}$ directly to
$\mbox{\boldmath{$\delta$}}_{lm\sigma}$.  Our analysis provides a simple
derivation of the expressions for the far-field flow resulting from a force
singularity in a slit pore.  The previously known formulas for the far-field
behavior of a Stokeslet \cite{Liron-Mochon:1976} and rotlet
\cite{Hackborn:1990} in a slit pore are special cases of our more general
expressions.

Considering Eq.\ \eqref{parameters of elements contributing to leading order}
and $m=\pm 1$ for parabolic flows in a slit pore, we find that only ${\bf
b}_{1\pm 10}$, ${\bf b}_{1\pm 11}$, ${\bf b}_{1\pm 12}$, ${\bf b}_{2\pm 10}$,
${\bf b}_{2\pm 11}$, and ${\bf b}_{3\pm 10}$ are the nonzero contributions in
Eq.\ \eqref{stat. av. bs}.  Moreover, as pointed out in Sec.\ \ref{Local
parallel-wall geometry}, we have further simplifications: $f_{1\pm 10}=f_{1\pm
11}=0$ (freely suspended particle) and $f_{lm\sigma}=f_{l\,-m\sigma}$ (choice
of local coordinates). Hence, only four different coefficients $f_{lm\sigma}$
($f_{112}$, $f_{210}$, $f_{211}$, $f_{310}$) and eight different source terms
${\bf b}_{lm\sigma}$ (${\bf b}_{1\pm 12}$, ${\bf b}_{2\pm 10}$, ${\bf b}_{2\pm
11}$, ${\bf b}_{3\pm 10}$) are involved in the expression for the pressure
source ${\bf b}_{s}$.  In Appendix \ref{derv. liron mochon} we show that these
source terms are given by the expressions
\begin{equation}
\label{specific blms}
{\bf b}_{l1\sigma}({\bf r},{\bf r}_0)={\bf b}_{l\,-1\sigma}^*({\bf r},{\bf r}_0)=d_{l\sigma}({\bf r}_0)(\hat{\bf e}_{\rho}+i\hat{\bf e}_{\phi})\delta(\mbox{\boldmath{$\rho$}}-\mbox{\boldmath{$\rho$}}_0),
\end{equation}
where the asterisk denotes the complex conjugate,
\mbox{$\delta(\brho-\brho_0)$} is the
Dirac delta function and
\begin{subequations}
\label{specific dlms}
\begin{equation}
\label{112}
d_{12}=-12\sqrt{\frac{2\pi}{3}},
\end{equation}
\begin{equation}
\label{210}
d_{20}=6\sqrt{\frac{\pi}{30}}(\bar{h}-2z_0),
\end{equation}
\begin{equation}
\label{211}
d_{21}=12\sqrt{\frac{\pi}{30}},
\end{equation}
\begin{equation}
\label{310}
d_{30}=-12\sqrt{\frac{4\pi}{4725}}.
\end{equation}
\end{subequations}
In the above relations, $\bar h$ denotes the local gap width, and $\hat{\bf
e}_{\rho}$ and $\hat{\bf e}_{\phi}$ represent the unit vectors along the
radial and azimuthal directions (note that these vectors correspond to the
basis vectors $\be_{x'}$ and $\be_{y'}$ in the local coordinate system
$x',y',z'$ centered on a small particle).

We combine the results \eqref{specific blms} and \eqref{specific dlms}
with Eq.\ \eqref{stat. av. bs} to find
\begin{equation}
\label{final bs}
{\bf b}_s(\mbox{\boldmath{$\rho$}})
   =12\bar{h}^3\hat{\bf e}_{\rho}\int\phi(\mbox{\boldmath{$\rho$}},z_0)
    d(\rho,z_0)dz_0,
\end{equation}
where $d$ is the dipolar strength of the far-field flow produced by a
small particle at the position $\rho,z_0$,
\begin{equation}
\label{d}
d(\rho,z_0)
   =\frac{b}{2\pi a \bar{h}^3}
    \sum_{l,\sigma}
        f_{l1\sigma}(\rho,z_0)
        d_{l\sigma}(\rho,z_0).
\end{equation}
This function is proportional to flow amplitude $A_L$ because
$f_{l1\sigma}$ varies linearly with $A_L$. Moreover, $d(\rho,z_0)$
depends on the radial coordinate $\rho$ only through the local gap
width $h(\rho)$ according to Eqs.\ \eqref{specific dlms} and the
assumption that the induced-force multipolar moments $f_{l1\sigma}$
are evaluated in the local parallel-wall-channel approximation.

\subsection{Numerical results for the pressure source}

\input{nidpl}

In Fig.\ \ref{nidpl-fig}, we present $d/A_L$ as a function of the
distance of the center of the small particle from the lower wall for
different gap widths where the position is normalized by the gap
width. In our calculations the results for the force multipolar
moments $f_{lm\sigma}$ were obtained using the multiple-scattering
method \cite{Bhattacharya-Blawzdziewicz:2002}, but the Cartesian-representation technique
\cite[][]{Bhattacharya-Blawzdziewicz-Wajnryb:2005, Bhattacharya-Blawzdziewicz-Wajnryb:2005a,
Bhattacharya-Blawzdziewicz-Wajnryb:2006a} yields equivalent results.
In Fig.\ \ref{nidpl-fig} we show only one half of the channel because
$d$ is an even function of the particle position with respect to the
channel mid-plane.
 
We find that the maximum contribution in $d$ comes from the multipole
corresponding to the induced stresslet ($l=2$, $\sigma=1$). When the
particle is close to the wall, the induced stresslet increases to a
large value because the slow inverse-logarithmic decay of the particle
mobility creates a large hydrodynamic stress in the lubrication
region. However, the dipolar strength $d$ always remains finite,
because for extremely small gaps the particle--wall relative velocity
ultimately decreases to zero.

There is no induced stresslet when the particle is at the center of
the channel. For this position, only higher order multipoles
contribute in $d$ and therefore $d$ has a minimum at the channel
center.
 
For large gap widths and intermediate positions of the
particle where it is neither close to the wall nor near the center of
the channel, the curves for different $h$ match with the curve defined
by
\begin{equation}
\label{d at infinity}
d=5\,\frac{A_{L}}{\bar{h}^3}\,(1-2z/\bar{h})^{2}.
\end{equation}
This can be predicted by calculating the strength of the induced
stresslet due to hydrodynamic interactions between a parabolic flow
and a sphere in an infinitely wide gap.

\section{Evaluation of the effective resistance force $F^{\rm c}_{0}$}
\label{Effective resistance force}

In this section, we focus on the evaluation of the correction $F^{\rm c}_{0}$
to the hydrodynamic resistance coefficient for the large particle moving near
the wall in a solution of small particles.  To this end we first determine the
pressure field for different ranges of the excluded shell $\bar\delta$.  The
resistance coefficient is then evaluated by integrating the pressure.

\input{dpl}

\input{pressure}

\input{force}

\subsection{Evaluation of the pressure field}
\label{Evaluation of the pressure field}

For a given $d$, we can determine the resistance force $F^{\rm c}_{0}$
by combining Eqs.\ \eqref{lubrication prs. eq.}, \eqref{leading order
force and pressure} and \eqref{final bs}.  Here we perform explicit
calculations for a model system of spheres interacting via the
excluded-shell potential described in Sec.\ \ref{Geometry}.  Since in
the short-time (high-frequency) limit the ensemble average in Eq.\
\refeq{final bs} is taken over the equilibrium particle distribution,
the volume fraction of small particles is constant in the accessible
region,
\begin{equation}
\label{uniform phi}
\phi({\bf r})=\phi\qquad\mbox{when}\qquad 1/2+\bar{\delta}<z<\bar{h}-1/2-\bar{\delta}
\end{equation} 
and otherwise $\phi=0$. 

Using this model and Eq.\ \eqref{final bs}, we find
\begin{equation}
\label{mod. stat. av. div bs}
{\bf b}_{s}(\mbox{\boldmath{$\rho$}})=12\,\phi\,A_{L}\,\bar{h}^3\,D\,\hat{\bf e}_{\rho},
\end{equation}
where
\begin{equation}
\label{mod. stat. av. div bs new}
D=\int_{\frac{1}{2}+\bar{\delta}}^{\bar{h}-\frac{1}{2}-\bar{\delta}} \frac{d({\bf r}_{0})}{A_L}dz_{0}.
\end{equation}
We refer to $D$ as the integrated dipolar strength.

In Fig.\ \ref{dipole-fig}, we show the magnitude of $D$ as a function of the
ratio between the gap width and the diameter of the small particle. Different
curves correspond to different values of $\bar{\delta}$. When $h/(2a)\le
1+2\bar{\delta}$ the $D$ is zero because there are no small particles in the
gap.  The magnitude of $D$ however, increases rapidly to a maximum value when
the gap is wide enough to accommodate the particles. After reaching the
maximum, the curves decay with the increasing gap because of the decreasing
strength of the induced force multipoles.

When a parabolic flow interacts with a sphere in a large gap, the
multipolar moments can be calculated without considering the wall
effects. In that case, the form \eqref{d at infinity} of the induced
stresslet in the local shear flow yields
\begin{equation}
\label{D at infty gap}
D=\frac{5}{3\bar{h}^{2}}.
\end{equation}
Figure \ref{dipole-fig} shows that the curves approach a slope of $-2$ in the log---log plot
for large $h$.

Because of radial symmetry, all associated quantities in Eq.\
\eqref{mod. stat. av. div bs} like $A_{L}$, $\bar{h}$ and $D$ are
functions of the radius $\rho$ only. Hence, ${\bf b}_{s}$ can be
expressed in terms of the gradient of a scalar
\begin{equation}
\label{mod. stat. av. div bs as gradient}
{\bf b}_{s}(\mbox{\boldmath{$\rho$}})=\bar{h}^3\bnabla S(\rho),
\end{equation}
where 
\begin{equation}
\label{the scalar s}
\frac{d\,S}{d\,\rho}=12\phi A_{L}D.
\end{equation}
Therefore, by substituting Eq.\ \eqref{mod. stat. av. div bs as gradient} in  Eq.\ \eqref{lubrication prs. eq.}, we find the perturbative pressure $p_0$:
\begin{equation}
\label{p eq s}
p_0=-S.
\end{equation} 
As a result, the radial derivative of $p_0$ can be obtained as a
radial function by combining Eqs.\ \eqref{the scalar s} and \eqref{p
eq s} along with the expression \eqref{al} for amplitude $A_{L}$ of
the incident parabolic lubrication flow
\begin{equation}
\label{radial derivative of pressure}
\frac{d\,p_{0}}{d\,\rho}= -9\phi\,\frac{\rho\,D}{\bar{h}}.
\end{equation}
We determine $p_{0}$ by integrating Eq.\ \eqref{radial derivative of pressure}.

A further simplification can be achieved by noting that in the
lubrication region the surface of the large particle is approximately
given by
\begin{equation}
\label{h and rho}
\bar{h}(\rho)=\bar{h}_{0}+\frac{\rho^{2}}{2}.
\end{equation}
Assuming \eqref{h and rho}, the pressure can be expressed as an
explicit function of the local gap width:
\begin{equation}
\label{gap derivative of pressure}
\frac{d\,p_{0}}{d\,\bar{h}}=-9\phi\,\frac{D}{\bar{h}}.
\end{equation}

In Fig.\ \ref{pressure-fig}, the perturbation pressure field is
plotted as a function of the gap width for different values of
$\bar{\delta}$. The plateau region of the curves corresponds to $D=0$
in the region depleted of small spheres because of the geometrical
constraints. With the increasing range of the excluded-volume
potential $\bar\delta$, the plateau region becomes wider, and a
smaller number of particles is accommodated in the lubrication
region. As a result $p_{0}$ decreases with increasing $\bar{\delta}$.

For a large gap, we find from Eq.\ \eqref{D at infty gap} that
\begin{equation}
\label{pressure at infty gap}
p_{0}=\frac{15\phi}{2\bar{h}^{2}}.
\end{equation}
which agrees with the result obtained using Einstein's formula for the
effective viscosity. This asymptotic value is independent of the range
of the excluded shell, provided that $\bar{\delta}\ll\bar h_0$. This
behavior is seen in Fig.\ \ref{pressure-fig} where for a large gap all
the curves coincide with an asymptotic line of a slope of $-2$ in the
log--log plot.

\subsection{Friction correction}
\label{Friction correction for the large particle}

We now use Eq.\ \eqref{leading order force and pressure} to obtain the
correction to the normal friction coefficient of the large particle
due to the presence of the smaller species. Owing to the axial
symmetry, the integral in Eq.\ \eqref{leading order force and
pressure} is reduced to the following form with the help of Eq.\
\eqref{h and rho}:
\begin{equation}
\label{leading order force and pressure mod}
F^{\rm c}_{0}=\frac{\pi\eta b^{2}}{a}\,\int_{\bar{h}_{0}}^{\infty}(p_{0}- p_{0}^{\infty})\,d\bar{h},
\end{equation}

The force $F^{\rm c}_{0}$, evaluated from Eq.\ \eqref{leading order force and
pressure mod}, is presented in Fig.\ \ref{force-fig} as a function of the
minimum separation $h_{0}$ between the wall and the surface of the large
particle.  The results are normalized by the product of the volume fraction
$\phi$, and the lubrication force \refeq{brenner} acting on the large particle
in the absence of the small particles.  The normalized correction to the
friction force decreases with increasing $\bar{\delta}$ because fewer spheres
are accommodated in the lubrication zone. For a larger gap, the ratio $F^{\rm
c}_{0}/F_{0}$ approaches a value of $2.5\phi$ for all values of
$\bar{\delta}$.  This is evident in Fig.\ \ref{force-fig} and consistent with
Einstein's formula for the effective viscosity of the dilute colloidal
solutions \cite[][]{Einstein:1906}.

\section{Conclusions}
\label{Conclusion}

This article presents a new lubrication theory for analyzing the near-wall
dynamics of a large colloidal particle in the presence of smaller
particles. Our theory is applied to determine the effect of the smaller
species on the hydrodynamic friction of the large particle when it moves in
the direction normal to the wall. This effect is quantified in terms of the
relative contribution $F^{\rm c}_{0}$ to short-time (or high-frequency) normal
friction force acting on the large particle.

We compute $F^{\rm c}_{0}$ as a function of the minimum separation between the
surface of a large particle and the wall. We model the screened electrostatic
repulsion between the solid surfaces in terms of excluded volumes and vary the
range of the excluded-volume potential to find its effect on the normal
friction. Our results show a decrease in friction with an increase in the
excluded volume as the number of smaller particles decreases in the
near-contact region.

The key simplification in our formulation stems from the fact that the
predominant hydrodynamic effect of the small particles is due to the far-field
form of the scattered flow from these particles. Our lubrication theory can be
applied to evaluate this far-field velocity (which is basically a
pressure-driven Hele--Shaw flow in a slit pore). Our analysis provides a
concise derivation of the general far-field solution for the flow produced by
an arbitrary force singularity between two parallel walls.  This derivation is
much simpler and more general than the earlier analyses of the Stokeslet
\cite[][]{Liron-Mochon:1976} and rotlet \cite[][]{Hackborn:1990} flow in a
slit pore geometry.

We have verified our theory by comparing our calculations with known limiting
results.  In particular, when the size of the smaller species is much smaller
than the separation between the large particle and the wall, the calculated
lubrication pressure field and normal friction correction agree with the
expressions obtained by using Einstein's effective viscosity formulation
\cite[][]{Einstein:1906}.

Our analysis can be extended to describe a more general near-contact
dynamics of bidisperse suspensions.  For example, our lubrication
theory can also be applied to cases where non-spherical particles and
non-planar walls are involved. In the future, we will focus on such
a generalization of the present analysis.  We will also describe the
effect of Brownian motion on the system dynamics
\cite{Bhattacharya:2005a}.

\acknowledgments{JB would like to acknowledge the support of this work by NSF
CAREER grant CTS-0348175.} 

\appendix

\section{Integral derivation of far-field flow in a slit pore}
\label{derv. liron mochon}

In this Appendix, we present the integral derivation of the far-field results
described in section 3. As a corollary of our derivation, we recover the
expressions for the well-known far-field flow solutions due to Stokeslet
\cite[][]{Liron-Mochon:1976} and rotlet \cite[][]{Hackborn:1990}.

First, we list the horizontal component of
$\mbox{\boldmath{$\delta$}}_{lm\sigma}$ defined in Eq.\ \eqref{single
sphere force distribution}, expressing the higher order multipoles as
derivatives of Stokeslets. For our purpose, we only have to consider
$m=\pm 1$ which corresponds to the parabolic flow. Moreover, the only
multipolar forces $\mbox{\boldmath{$\delta$}}_{lm\sigma}$ that are
relevant are those which give non-zero ${\bf b}_{lm\sigma}$ in Eq.\
\eqref{b l m sigma}. Hence, we take Eq.\ \eqref{parameters of elements
contributing to leading order} into account and only analyze the cases
where $l+\sigma<3$.

For $m=1$, the horizontal component of the relevant
$\mbox{\boldmath{$\delta$}}_{lm\sigma}$ is given below:
\begin{subequations}
\label{deltas}
\begin{equation}
\label{delta 1}
\mbox{\boldmath{$\delta$}}^{\parallel}_{110}({\bf r})=-\sqrt{\frac{2\pi}{3}}\,\delta({\bf r})\,\be,
\end{equation}
\begin{equation}
\label{delta 2}
\mbox{\boldmath{$\delta$}}^{\parallel}_{111}({\bf r})=\sqrt{\frac{2\pi}{3}}\,\frac{\partial\delta}{\partial z}\,\be,
\end{equation}
\begin{equation}
\label{delta 3}
\mbox{\boldmath{$\delta$}}^{\parallel}_{112}({\bf r})=\sqrt{\frac{2\pi}{3}}\,\frac{\partial^{2}\delta}{\partial z^{2}}\,\be,
\end{equation}
\begin{equation}
\label{delta 4}
\mbox{\boldmath{$\delta$}}^{\parallel}_{210}({\bf r})=-\sqrt{\frac{\pi}{30}}\,\frac{\partial\delta}{\partial z}\,\be,
\end{equation}
\begin{equation}
\label{delta 5}
\mbox{\boldmath{$\delta$}}^{\parallel}_{211}({\bf r})=-\sqrt{\frac{\pi}{30}}\,\frac{\partial^{2}\delta}{\partial z^{2}}\,\be,
\end{equation}
\begin{equation}
\label{delta 6}
\mbox{\boldmath{$\delta$}}^{\parallel}_{310}({\bf r})=\sqrt{\frac{4\pi}{4725}}\,\frac{\partial^{2}\delta}{\partial z^{2}}\,\be,
\end{equation}
\end{subequations}
where $\delta({\bf r})$ is the Dirac-delta function and the superscript $\parallel$ denotes the projection of the vector on $x$-$y$ plane. The vector $\be$ is defined as
\begin{equation}
\label{e vector}
\be=\hat{\bf e}_{x}+{\rm i}\hat{\bf e}_{y}
\end{equation}
(except for normalization \be is identical to ${\bf e}_{+1}$ defined in \cite{Edmonds:1960}). 
For $m=-1$, the source terms $\mbox{\boldmath{$\delta$}}_{lm\sigma}$ are the complex conjugates of
$\mbox{\boldmath{$\delta$}}_{l1\sigma}$. Therefore, we only focus on
the force distribution described in Eq.\ \eqref{deltas}.

In the above expressions for $\mbox{\boldmath{$\delta$}}_{lm\sigma}$
only $\delta$, and the derivatives $\partial\delta/\partial z$ and
$\partial^{2}\delta/\partial z^{2}$ appear in
addition to the combinatorial coefficients. Hence, in order to obtain
${\bf b}_{lm\sigma}$ by using Eqs.\ \eqref{b l m sigma} and
\eqref{deltas} we derive the following relations by evaluating the
integrals in the definition of operator $\hat{\sf B}$ in Eq.\ \eqref{B
operator}:
\begin{subequations}
\label{B delta}
\begin{eqnarray}
&&\hat{\sf B}\,\delta({\bf r}-{\bf r}_0)=6 z_0(\bar{h}-z_0)\delta(\mbox{\boldmath{$\rho$}}-\mbox{\boldmath{$\rho$}}_0)\qquad\\
&&\hat{\sf B}\,\frac{\partial}{\partial z}\delta({\bf r}-{\bf r}_0)=-6(\bar{h}-2z_0)\delta(\mbox{\boldmath{$\rho$}}-\mbox{\boldmath{$\rho$}}_0)\qquad\\
&&\hat{\sf B}\,\frac{\partial^{2}}{\partial z^{2}}\delta({\bf r}-{\bf r}_0)=-12\delta(\mbox{\boldmath{$\rho$}}-\mbox{\boldmath{$\rho$}}_0).
\end{eqnarray}
\end{subequations}
By combining relations \eqref{deltas} and \eqref{B delta} with Eq.\ \eqref{b l
m sigma} we find 
\begin{equation}
\label{specific blms ap}
{\bf b}_{l1\sigma}({\bf r},{\bf r}_0)={\bf b}_{l\,-1\sigma}^*({\bf r},{\bf r}_0)=d_{l\sigma}({\bf r}_0)\be\delta(\mbox{\boldmath{$\rho$}}-\mbox{\boldmath{$\rho$}}_0),
\end{equation}
where
\begin{subequations}
\label{specific dlms ap}
\begin{equation}
\label{110 ap}
d_{10}=-2\sqrt{6\pi}z_0(\bar{h}-z_0),
\end{equation}
\begin{equation}
\label{111 ap}
d_{11}=-2\sqrt{6\pi}(\bar{h}-2z_0),
\end{equation}
\end{subequations}
and the remaining nonzero coefficients are given by Eq.\ \eqref{specific dlms}.
The relations \eqref{110 ap} and \eqref{111 ap} are associated with the
far-field Hele--Shaw flow produced by a Stokeslet and a rotlet in
parallel-plate geometry. Resulting expressions for $b_{110}$ and $b_{111}$ are
in agreement with Liron--Mochon's \cite{Liron-Mochon:1976} and Hackborn's
\cite{Hackborn:1990} expressions for the far-field solutions whereas the other $b_{lm\sigma}$ 
described in the article are equivalant to the coefficients derived in 
our earlier paper \cite[][]{Bhattacharya-Blawzdziewicz-Wajnryb:2006} by using a more complicated matrix representation.

\bibliographystyle{unsrt}
\bibliography{sbib}
%\begin{widetext}
%\listoffigures 
%\end{widetext}
\end{document}

%% file: geo.tex
\begin{figure*}
\begin{center}

\begin{picture}(470,220)(0,0)
\put(50,-10){\scalebox{1.0}{
\begin{picture}(0,0)(0,0)
    \put(0,0){\scalebox{0.60}{\includegraphics
       {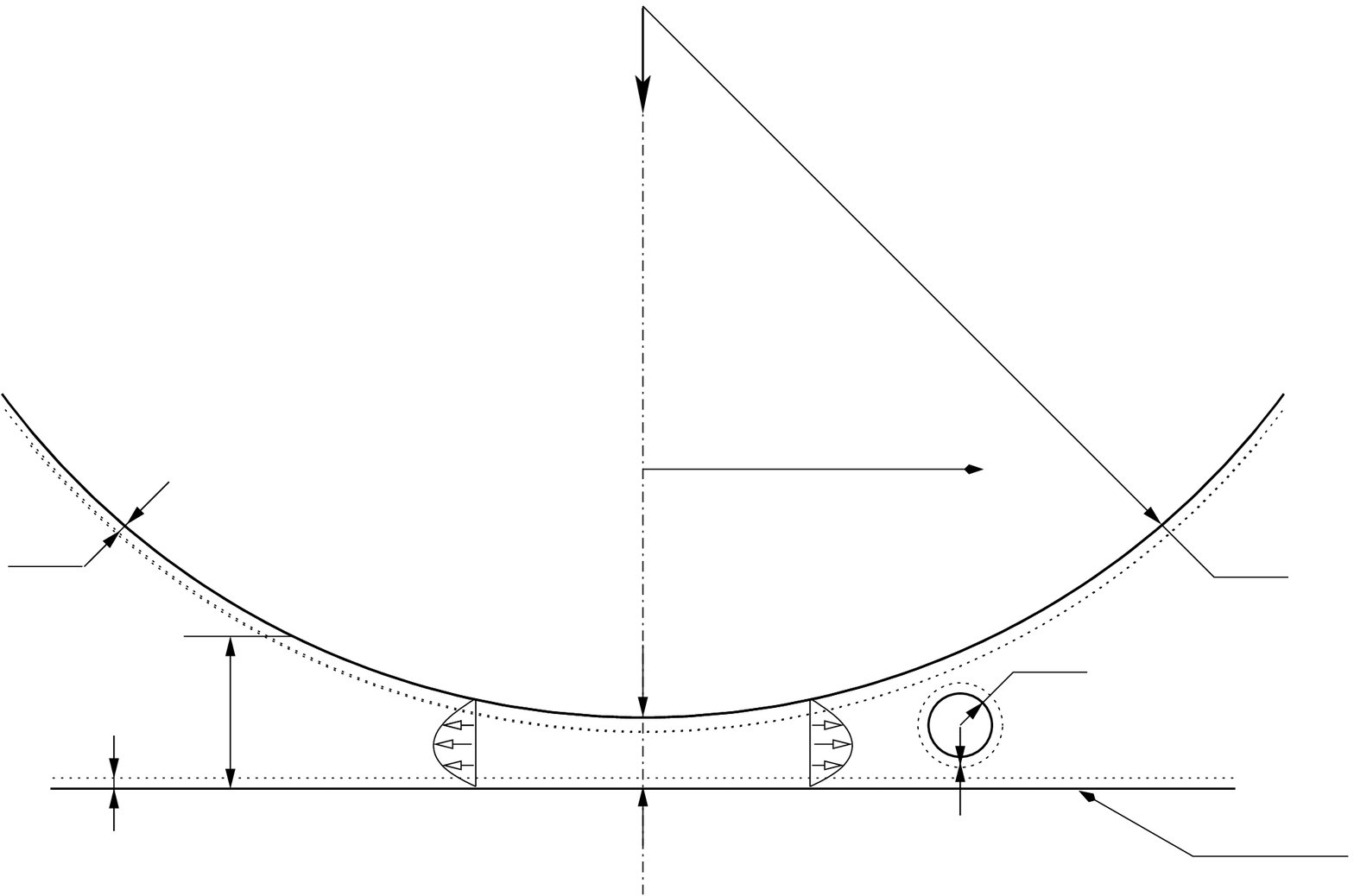}}}
  \end{picture}
\put(10,87){$\delta_{b}$}
\put(18,18){$\delta_{w}$}
\put(237,18){$\delta_{a}$}
\put(265,60){$a$}
\put(316,85){$b$}
\put(40,47){$h(\rho)$}
\put(154,34){$h_{0}$}
\put(150,200){$V^{\rm P}$}
\put(225,112){$\rho$}
\put(312,12){$z=0$}
}}
\end{picture}

\end{center}
\input{geo-cap}
\end{figure*}

%% file: geo-cap.tex
\caption{
Near-contact configuration of a large particle in the presence of a
small particle in a bidisperse colloidal solution. Dotted lines
represent the boundary of the excluded volumes due to electrostatic
repulsion.
}

\label{geo-fig}

%% file: nidpl.tex
\begin{figure}

\begin{picture}(0,205)(200,0)

\put(72,5){\scalebox{0.7}{\includegraphics{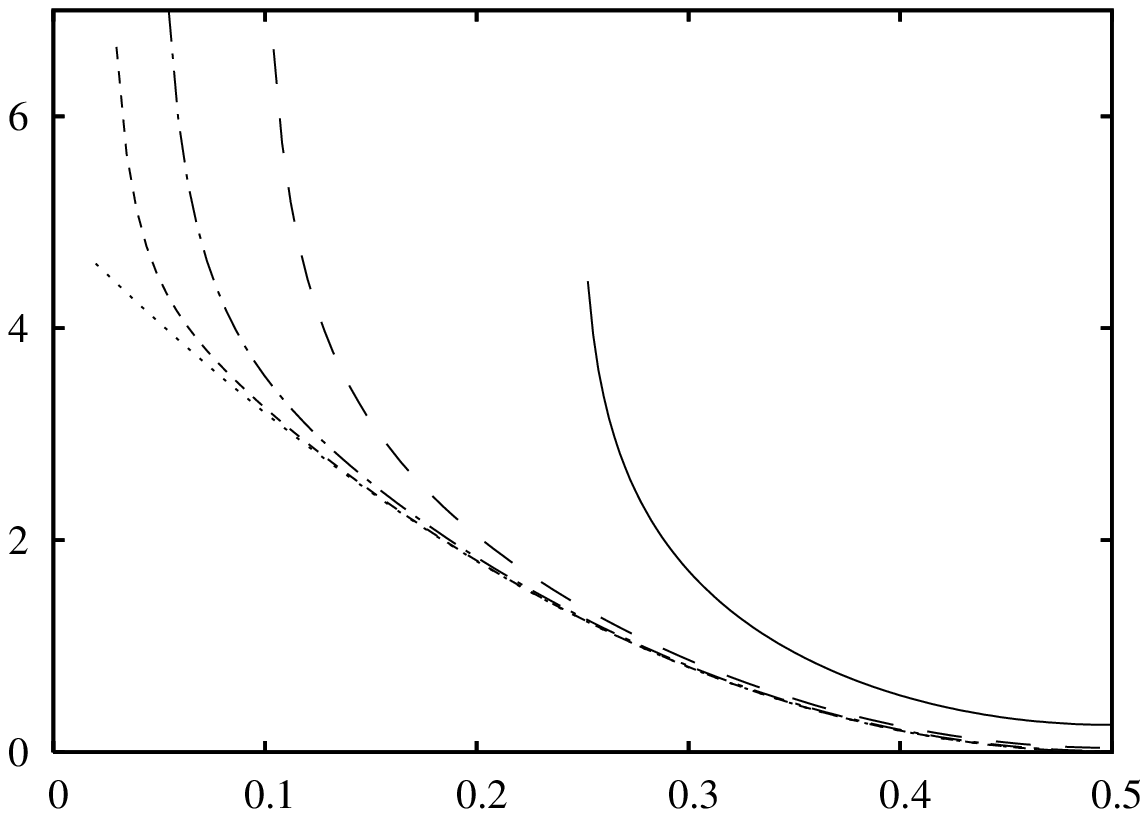}}}
\put(69,88){$\displaystyle\frac{\bar{h}^3}{A_{l}}d$}
\put(185,4){$z/\bar{h}$}

\end{picture}
\input{nidpl-cap}
\end{figure}

%% file: nidpl-cap.tex
\caption{
Normalized induced-dipole strength for two-dimensional Hele--Shaw
pressure field as a function of position of the small-particle center
from the wall for $\bar h=2.0$ (solid line), $\bar{h}=5.0$ (long
dashed line), $\bar{h}=10.0$ (dash-dot line) and $\bar{h}=20.0$ (short
dash line). The dotted line represents the exact result for a very
small particle in a parabolic flow when the wall is situated far away
from the sphere.
}

\label{nidpl-fig}

%% file: dpl.tex
\begin{figure}

\begin{picture}(0,205)(200,0)

\put(65,5){\scalebox{0.7}{\includegraphics{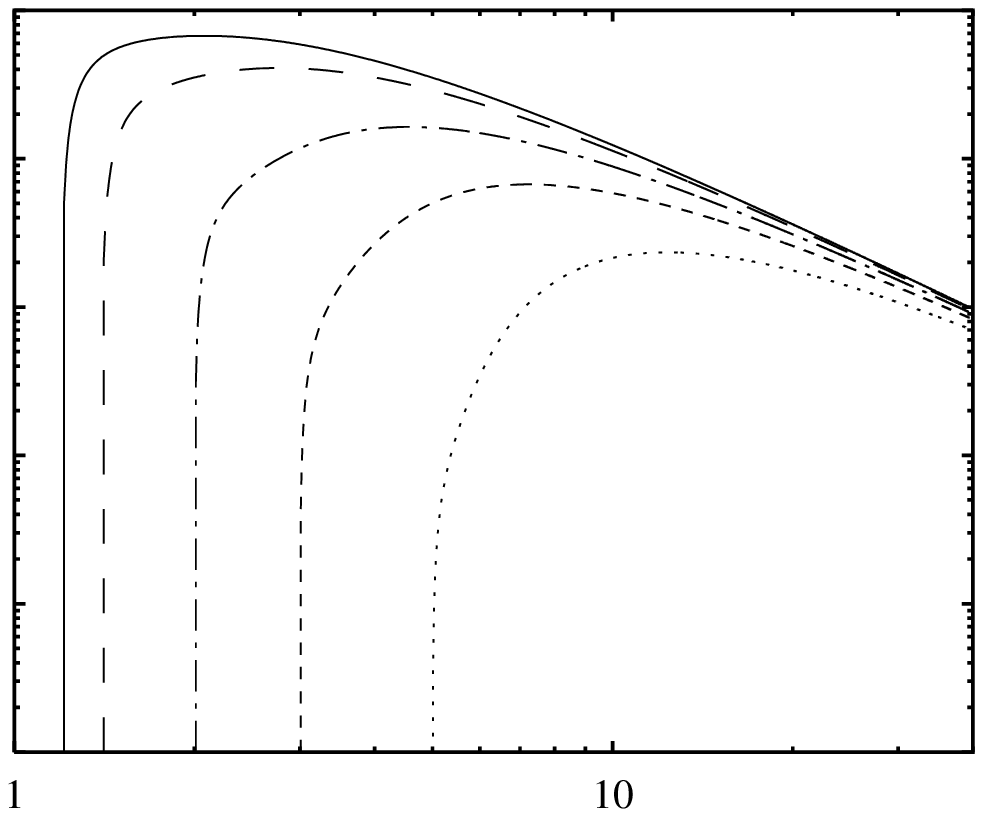}}}
\put(88,18){$10^{-6}$}
%\put(88,46){$10^{-5}$}
\put(88,78){$10^{-4}$}
%\put(88,106){$10^{-3}$}
\put(88,138){$10^{-2}$}
%\put(88,166){$10^{-1}$}
\put(90,96){$D$}
\put(185,4){$h(\rho)/2a$}

\end{picture}
\input{dpl-cap}
\end{figure}

%% file: dpl-cap.tex
\caption{
Normalized integrated dipolar strength for two-dimensional Hele--Shaw pressure
field as a function of normalized local gap for $\bar{\delta}=0.1$ (solid
line), $\bar{\delta}=0.2$ (long dashed line), $\bar{\delta}=0.5$ (dash-dot
line), $\bar{\delta}=1.0$ (short dash line) and $\bar{\delta}=2.0$ (dotted
line).
}

\label{dipole-fig}

%% file: pressure.tex
\begin{figure}

\begin{picture}(0,205)(200,0)

\put(65,5){\scalebox{0.7}{\includegraphics{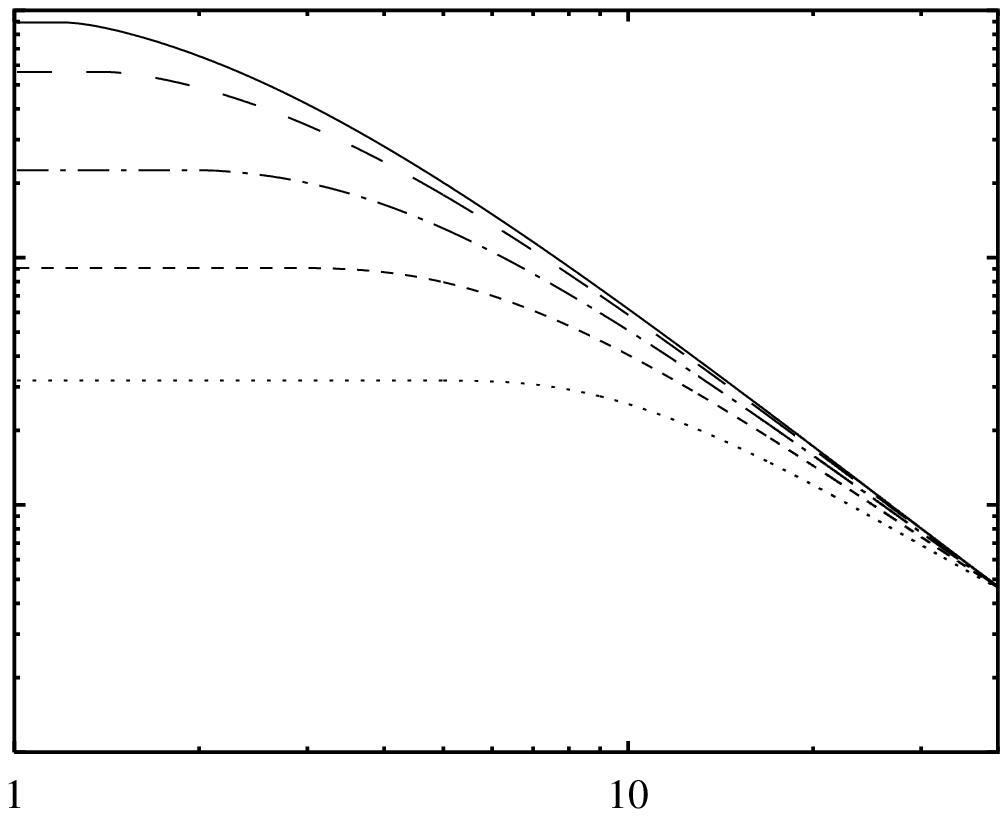}}}
\put(83,18){$10^{-3}$}
\put(83,68){$10^{-2}$}
\put(83,118){$10^{-1}$}
\put(88,168){$10^{0}$}
\put(85,96){$\displaystyle\frac{p_{0}}{\phi}$}
\put(185,4){$h(\rho)/2a$}

\end{picture}
\input{pressure-cap}
\end{figure}

%% file: pressure-cap.tex
\caption{
Correction to the lubrication pressure in the near-contact region due
to the presence of small particles, normalized by $6\pi\eta b
v_{P}/a^2$, vs. the ratio between the local gap width and the
small-particle radius, for different values of $\bar\delta$ (indicated
by the same line types as in Fig.\ref{dipole-fig}).
}

\label{pressure-fig}

%% file: force.tex
\begin{figure}

\begin{picture}(0,205)(192,0)

\put(65,5){\scalebox{0.7}{\includegraphics{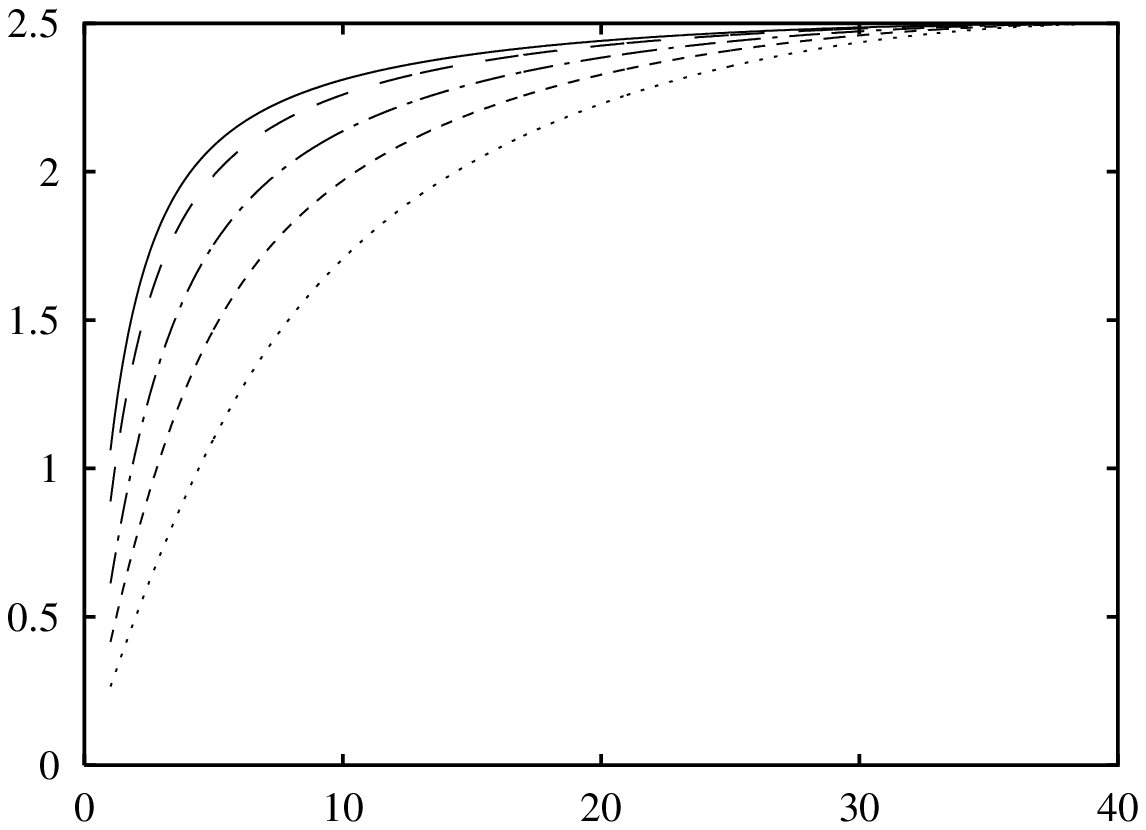}}}
\put(61,94){$\displaystyle\frac{F^{\rm c}_{0}}{\phi F_{0}}$}
\put(186,0){$h_{0}/2a$}

\end{picture}
\input{force-cap}
\end{figure}

%% file: force-cap.tex
\caption{
Small-particle correction to the friction force acting on the large
particle, normalized by $F_{0}=6\pi\eta b^{2}/h_{0} $, vs. the ratio
between the minimum gap width at $\rho=0$ and the small particle
radius, for different values of $\bar\delta$ (indicated by the same
line types as in Fig.\ref{dipole-fig}).
}

\label{force-fig}